\documentstyle[12pt]{article}

\begin{document}
\title{Generalized Weyl-Wigner map and Vey quantum mechanics}
\author{
Nuno Costa Dias\footnote{{\it mop48451@mail.telepac.pt}} \\ { {\it Grupo de Astrof\'{\i}sica e Cosmologia - Departamento de F\'{\i}sica}} \\ {{\it Universidade da Beira Interior, 6200 Covilh\~{a}, Portugal}} \\ 
\\
Jo\~ao Nuno Prata\footnote{{\it mat@ulusofona.pt}} \\ {{\it 
Universidade Lus\'ofona de Humanidades e Tecnologias}} \\ {{\it Av. Campo Grande, 376,  
1749-024 Lisboa, Portugal}}}
\maketitle

\begin{abstract}
The Weyl-Wigner map yields the entire structure of Moyal quantum mechanics directly from the standard operator formulation. The covariant generalization of Moyal theory, also known as Vey quantum mechanics, was presented in the literature many years ago. However, a derivation of the formalism directly from standard operator quantum mechanics, clarifying the relation between the two formulations is still missing. In this paper we present a covariant generalization of the Weyl order prescription and of the Weyl-Wigner map and use them to derive Vey quantum mechanics directly from the standard operator formulation. The procedure displays some interesting features: it yields all the key ingredients and provides a more straightforward interpretation of the Vey theory including a direct implementation of unitary operator transformations as phase space coordinate transformations in the Vey idiom. These features are illustrated through a simple example.    
\end{abstract}

\section{Introduction}

The Weyl-Wigner map \cite{Weyl,Wigner} yields the Moyal formulation of quantum mechanics, \cite{Moyal,Wigner2,Lee1,Carruthers,Balazs,Fairlie1} alternative to the more conventional standard operator \cite{Bohr,Heisenberg,Dirac,Cohen} and path integral formulations. The main features of Moyal quantum mechanics are that it is formulated in terms of phase space functions and dynamics is based on a deformation of the Poisson bracket, named Moyal bracket \cite{Moyal,Baker,Flato1,Flato2}.  
This formulation of quantum mechanics has been receiving increased attention namely in the context of the fields of the semiclassical limit of quantum mechanics \cite{Lee1,Flato1,Flato2,Lee2,Smith,nuno1,nuno3}, quantum chaos \cite{latka,shin}, hybrid dynamics \cite{nuno2,nuno4} and also in $M$-theory \cite{Fairlie2,Fairlie3}.

The Weyl-Wigner isomorphism between operators and phase space functions (symbols) provides the entire structure of Moyal quantum mechanics directly from the standard operator formulation. 
Let us choose a set of fundamental operators $(\hat q_i,\hat p_i )$ and the corresponding set of canonical variables $(q_i,p_i), i=1..N$ for an arbitrary $N$ dimensional dynamical system. The Weyl-Wigner map $W_{(q,p)}: {\hat{\cal A}} \to {\cal A}(T^* M)$ attributes to a given operator $\hat A$ in the quantum algebra $ {\hat{\cal A}}$ a unique element of the algebra of functions over the phase space $T^* M$: 
\begin{equation}
W_{(q,p)}(\hat A)= \int d^N\vec y e^{-i\vec p \cdot \vec y} <\vec q-\frac{\hbar}{2} \vec y|\hat A|\vec q+\frac{\hbar}{2} \vec y>,
\end{equation}
where we used the compact notation:
$$
\vec y \equiv (y_1, \cdots, y_N);
\quad 
d^N \vec y \equiv dy_1 \cdots dy_N;
\quad 
{\vec p} \equiv (p_1, \cdots ,p_N)
\quad
{\vec q} \equiv ( q_1, \cdots , q_N),
$$
and the subscript $(q,p)$ means that the corresponding object (the Weyl-Wigner map in this case) is defined in the variables $(\vec q,\vec p)$. This specification seems redundant now but is important for the sequel.
The Weyl-Wigner map is bijective and univocous. Moreover, it is an isomorphism between the quantum $\left({\hat{\cal A}}, \cdot , \left[, \right] \right)$
and the "classical" $\left( {\cal A} , * , \left[, \right]_M \right)$ algebras. The quantum algebra ${\hat{\cal A}}$ is based on the operator product
$\cdot$ and the quantum commutator $\left[, \right]$, whereas the classical algebraic structures are the "star" product $*$ and the Moyal bracket 
$\left[, \right]_M$. The Weyl-Wigner map is a morphism in the sense that:
\begin{equation}
\begin{array}{l l}
W_{(q,p)} (\hat A \cdot \hat B) & =  W_{(q,p)} (\hat A) *_{(q,p)} W_{(q,p)} (\hat B),\\
& \\
W_{(q,p)} \left( \left[ \hat A, \hat B \right] \right) & = \left[ W_{(q,p)} (\hat A ), W_{(q,p)} (\hat B) \right]_{M_{(q,p)}}, \qquad \forall_{\hat A, \hat B \in {\hat{\cal A}}},
\end{array}
\end{equation}
and it yields the functional structure of the star product and Moyal bracket: 
\begin{equation}
A*_{(q,p)} B = A e^{\frac{i \hbar}{2} {\hat{\cal J}}_{(q,p)}} B, \qquad \left[ A, B \right]_{M_{(q,p)}} = \frac{2}{\hbar}  A \sin \left(\frac{\hbar}{2} {\hat{\cal J}}_{(q,p)} \right) B,
\qquad A, B \in {\cal A}
\end{equation}
where ${\hat{\cal J}}_{(q,p)}$ is the "{\it Poisson}" operator: $
{\hat{\cal J}}_{(q,p)} \equiv \sum_{i=1}^N \left( \frac{ {\buildrel { \leftarrow}\over\partial}}{\partial q_i} \frac{ {\buildrel { 
\rightarrow}\over\partial}}{\partial p_i} -  \frac{{\buildrel { \leftarrow}\over\partial}}{\partial p_i}  \frac{{\buildrel { 
\rightarrow}\over\partial}}{\partial q_i} \right)  $, 
the derivatives ${\buildrel { \leftarrow}\over\partial}$ and ${\buildrel { \rightarrow}\over\partial}$ acting on $A$ and $B$, respectively. Alternatively, 
${\hat{\cal J}}_{(q,p)}$ can be written as: ${\hat{\cal J}}_{(q,p)}={\buildrel { \leftarrow}\over\partial}_k J_{(q,p)}^{kl} {\buildrel { \rightarrow}\over\partial}_l$, where  
 $J_{(q,p)}^{kl}$ is the $kl$-th element of the symplectic matrix in the variables $(\vec q,\vec p)$:
\begin{equation}
J_{(q,p)} = \left( 
\begin{array}{l r}
0_{N \times N} & - 1_{N \times N}\\
1_{N \times N} & 0_{N \times N}
\end{array}
\right).
\end{equation}
We introduced the compact notation: $O^k = p_k,  k=1, \cdots ,N$; $O^k = q_{k-N},  k=N+1, \cdots, 2N$; ${\partial} / {\partial O_k} = \partial_k$ and sum over repeated indices is understood.

From eq.(3) it is trivial to obtain the following expansion in powers of $\hbar$:
\begin{equation}
A*_{(q,p)} B =  A \cdot B + \frac{i \hbar}{2}  A {\buildrel { \leftarrow}\over\partial}_k J_{(q,p)}^{kl} {\buildrel { \rightarrow}\over\partial}_l B 
+ \frac{1}{2} \left(  \frac{i \hbar}{2} \right)^2
 A {\buildrel { \leftarrow}\over\partial}_k {\buildrel { \leftarrow}\over\partial}_s 
J_{(q,p)}^{kl}J_{(q,p)}^{sn} 
{\buildrel { \rightarrow}\over\partial}_l {\buildrel { \rightarrow}\over\partial}_n B + \cdots
 \end{equation}
The Weyl-Wigner transform of the density matrix operator is the Wigner distribution function \cite{Wigner}, $f_W(\vec q,\vec p;t)=W_{(q,p)}(|\psi(t)><\psi(t)|)$, which is the fundamental mathematical object of Moyal quantum mechanics. Its time evolution is given by the dynamical equation:
\begin{equation}
\dot{f}_W(\vec q,\vec p;t)=[H(\vec q,\vec p),f_W(\vec q,\vec p;t)], \quad H=W_{(q,p)}(\hat H),
\end{equation}
where $\hat H$ is the quantum Hamiltonian.
We see that the mathematical structure of Moyal quantum mechanics is very similar to that of classical statistical mechanics. However, these similarities should not be taken too seriously. The procedure by which physical relevant information is obtained is a lot more elaborate. In classical statistical mechanics the fundamental predictions 
are the probabilities for finding the system in an arbitrary configuration $(\vec q_0,\vec p_0)$, which are given by the values of a true probability distribution function $\rho (\vec q=\vec q_0,\vec p=\vec p_0)$.

On the contrary, in Moyal quantum mechanics the value of $f_W (\vec q=\vec q_0,\vec p=\vec p_0)$  
cannot be given such straightforward interpretation, given the fact that $f_W(\vec q,\vec p)$ might take on negative values. The fundamental physical predictions of Moyal quantum mechanics are obtained through a procedure analogous to that of standard operator quantum mechanics. Given a general observable $A(\vec q,\vec p)$ we should solve the star-genvalue equation \cite{Fairlie1,nuno5}:
\begin{equation}
A(\vec q,\vec p) *_{(q,p)} g^n_a(\vec q,\vec p) = a g^n_a(\vec q,\vec p),
\end{equation}
where $n$ is a degeneracy index, to obtain the probability for a measurement of $A$ at the time $t$ yielding the value $a$:
\begin{equation}
P(A(\vec q,\vec p;t)=a)=\sum_{n} \int d^N\vec q \int d^N\vec p g^n_a(\vec q,\vec p) f_W(\vec q,\vec p;t),
\end{equation}
where we assumed that the degeneracy index is discrete. If this is not the case then the sum in $n$ is replaced by a (set of) integral(s) in $n$.
For the fundamental observables $q_i$ and $p_i$, $i=1..N$, eq.(8) reduces to a more "classical like" result:
\begin{equation}
P(q_i(t)=q_0)= \int d^N\vec q \int d^N\vec p  f_W(\vec q,\vec p;t) \delta(q_i-q_0); \quad P(p_i(t)=p_0)= \int d^N\vec q \int d^N\vec p  f_W(\vec q,\vec p;t) \delta(p_i-p_0),
\end{equation}
the same happening to the average value of $A(\vec q,\vec p)$:
\begin{equation}
<A(\vec q,\vec p;t)>= \int d^N\vec q \int d^N\vec p A(\vec q,\vec p) f_W(\vec q,\vec p;t).
\end{equation}

An important subject in any dynamical theory is the study of its invariances. Just like standard operator quantum mechanics the Moyal formulation is invariant under the action of general unitary transformations and contrary to what happens in classical mechanics is not invariant under the action of a significative set of coordinate transformations. In fact, most unitary operator transformations are not implemented as phase space coordinate transformations in the Moyal idiom.

Many years ago Vey \cite{Vey} presented a generalization of the star product that renders Moyal quantum mechanics fully invariant under phase space coordinate transformations. The original motivation was not to enlarge the invariance properties of Moyal quantum mechanics, but to derive the general form of the Poisson algebra deformations for curved phase spaces. Vey's original developments have been used in investigations aiming at two major directions. Firstly, in more mathematically oriented research, to generalize the Moyal-Weyl-Wigner quantization procedure to non-flat phase space manifolds \cite{Flato1,Flato2,Wilde,Omori}. Secondly, to provide a consistent classical interpretation of Moyal dynamics \cite{Flato1,Flato2}.

To our knowledge however, a complete study of the relation between Vey covariant quantum mechanics and standard operator quantum mechanics casting Vey theory at the same level of completeness as Moyal quantum mechanics has not yet been presented. Take for instance the Weyl-Wigner map. Although trivial, the covariant generalization of this map is still missing in the literature.

In this paper we emphasize the relation between standard operator quantum mechanics and Moyal quantum mechanics and attempt to derive the Vey formulation in a similar fashion. There are two main virtues in this approach: firstly it clarifies the relation between the standard operator and Vey quantum mechanics (providing, for instance a new point of view for the analysis
of the invariance properties of covariant phase space quantum mechanics).
Secondly it yields a (previously missing) covariant generalization of some key ingredients of phase space quantum mechanics, such as the Weyl-Wigner map, the Weyl order prescription, the average and the marginal probability distribution functionals and the star-genvalue equation.

This paper is organized as follows: in section 2 we start by reviewing some well known properties concerning the action of canonical and coordinate transformations in the Moyal formalism. We then present (section 3) a covariant generalization of the Weyl-Wigner map. The new map makes it possible to implement a general unitary transformation of standard operator quantum mechanics as a coordinate transformation in the Moyal idiom. In section 4 we use the new map to derive the covariant star product and, as a by-product, the dynamical structure of Vey quantum mechanics. In section 5 we present a summary of the structure of covariant quantum mechanics, including the generalizations of the star-genvalue equation, the average value and the probability functionals. Finally, in section 6 some of the features of the formalism are illustrated through a simple example.

Before proceeding let us make an important remark: from the outset we shall restrict our attention to the simpler case of dynamical systems displaying a phase space with the structure of a flat manifold.   

\section{Canonical and coordinate transformations} 

In classical mechanics all canonical transformations are phase space coordinate transformations. Their action is of the form (we shall take the passive point of view):
\begin{equation}
T:T^{\ast}M \longrightarrow T^{\ast}M; (\vec q,\vec p) \longrightarrow (\vec q=\vec q(\vec Q,\vec P),\vec p=\vec p(\vec Q,\vec P)),
\end{equation}
where $(\vec Q,\vec P)$ and $(\vec q,\vec p)$ are two sets of canonical variables. $T$
yields a transformation of a general observable given by:
\begin{equation}
T:A(\vec q,\vec p) \longrightarrow A^{\prime} (\vec Q,\vec P)=A (\vec q(\vec Q,\vec P),\vec p(\vec Q,\vec P)),  
\end{equation}
and for two general observables $A^{\prime}(\vec Q,\vec P)=T\{A(\vec q,\vec p)\}$ and 
$B^{\prime}(\vec Q,\vec P)=T\{B(\vec q,\vec p)\}$ we have:
\begin{equation}
\{A(\vec q(\vec Q,\vec P),\vec p(\vec Q,\vec P)),B(\vec q(\vec Q,\vec P),\vec p(\vec Q,\vec P))\}_{(q,p)} = \{A^{\prime}(\vec Q,\vec P),B^{\prime}(\vec Q,\vec P)\}_{(Q,P)},
\end{equation}
and thus the Hamiltonian equations of motion in the variables $(\vec q,\vec p)$ and $(\vec Q,\vec P)$ are fully equivalent: they yield identical mathematical solutions and thus identical physical predictions.

This picture does not translate to Moyal quantum mechanics. To see this explicitly let us go back to standard operator quantum mechanics and consider, to make it simpler, the unitary transformation:
\begin{equation}
\hat{\vec q} \equiv \hat{\vec q}(\hat{\vec Q},\hat{\vec P})=\hat U \hat{\vec Q} \hat U^{-1} \quad , 
\hat{\vec p} \equiv \hat{\vec p}(\hat{\vec Q},\hat{\vec P})=\hat U \hat{\vec P} \hat U^{-1} \quad ,
\hat A(\hat{\vec q},\hat{\vec p}) \equiv \hat U \hat A(\hat{\vec Q},\hat{\vec P}) \hat U^{-1} = \hat{A}'(\hat{\vec Q},\hat{\vec P}).
\end{equation}
The two sets of fundamental variables provide two Weyl-Wigner maps:
\begin{eqnarray}
W_{(Q,P)}(\hat A) &=& \int d^N\vec Y e^{-i\vec P\cdot \vec Y} <\vec Q-\frac{\hbar}{2}\vec Y|\hat A|\vec Q+\frac{\hbar}{2}\vec Y>,\nonumber \\
W_{(q,p)}(\hat A) &=& \int d^N\vec y e^{-i\vec p\cdot \vec y} <\vec q-\frac{\hbar}{2}\vec y|\hat A|\vec q+\frac{\hbar}{2}\vec y>,
\end{eqnarray}
from which we can derive the action of unitary transformations in the Moyal formalism. The fundamental variables transform trivially: 
\begin{equation}
T:(\vec q,\vec p)\longrightarrow \left( \vec q=W_{(Q,P)}(\hat{\vec q}(\hat{\vec Q},\hat{\vec P}))=\vec q(\vec Q,\vec P), 
\vec p=W_{(Q,P)}(\hat{\vec p}(\hat{\vec Q},\hat{\vec P}))=\vec p(\vec Q,\vec P)\right) ,
\end{equation}
and a general observable transforms as:
\begin{eqnarray}
T:A(\vec q,\vec p) = W_{(q,p)} (\hat A(\hat{\vec q},\hat{\vec p})) \longrightarrow  A^{\prime}(\vec Q,\vec P) & = &  
W_{(Q,P)} (\hat{A}'(\hat{\vec Q},\hat{\vec P})) \nonumber \\
& = & U(\vec Q,\vec P) *_{(Q,P)} A(\vec Q,\vec P)*_{(Q,P)} U^{-1}(\vec Q,\vec P),  
\end{eqnarray}
which, in general does not correspond to the action of a coordinate transformation (except if $T$ is linear) since:
\begin{equation}
A^{\prime}(\vec Q,\vec P) \not= A(\vec q(\vec Q,\vec P),\vec p(\vec Q,\vec P)),
\end{equation}
even though the transformation is canonical. For two general observables $A(\vec q,\vec p)$ and $B(\vec q,\vec p)$ we have:
\begin{equation}
[A'(\vec Q,\vec P),B'(\vec Q,\vec P)]_{M_{(Q,P)}} = T([A(\vec q,\vec p),B(\vec q,\vec p)]_{M_{(q,p)}}),
\end{equation}
and thus the Moyal dynamical equations in the variables $(\vec Q,\vec P)$ and $(\vec q,\vec p)$:
\begin{equation}
\dot{A}(\vec q,\vec p;t)=[A(\vec q,\vec p;t),H(\vec q,\vec p)]_{M_{(q,p)}}
\quad \mbox{and} \quad   
\dot{A}'(\vec Q,\vec P;t)=[A'(\vec Q,\vec P;t),H'(\vec Q,\vec P)]_{M_{(Q,P)}} ,
\end{equation}
yield two mathematical solutions, related by:
\begin{equation}
A(t)=F(\vec q,\vec p,t) \quad \mbox{and} \quad A^{\prime} (t) = F^{\prime}(\vec Q,\vec P,t)=U*_{(Q,P)}F(\vec Q,\vec P,t)*_{(Q,P)}U^{-1},
\end{equation}
which, in general are not the same phase space function: $
F^{\prime}(\vec Q,\vec P,t) \not= F(\vec q(\vec Q,\vec P),\vec p(\vec Q,\vec P),t)$, albeit providing the same physical predictions: eqs.(8,9,10).
We see that no physical meaning can be attached to a single value of the observable $A$ since this value is dependent of the particular representation chosen. Take for instance the Wigner distribution function that may be positive defined in one representation and become negative under a unitary transformation. 

On the other hand, most coordinate transformations act non-canonically in the Moyal formalism (the exceptions, once again, are linear transformations): consider the transformation $T$ eq.(16) and the two general phase space functions $G(\vec q,\vec p)$ and $F(\vec q,\vec p)$. In general we have:
\begin{equation}
[G(\vec q(\vec Q,\vec P),\vec p(\vec Q,\vec P)),F(\vec q(\vec Q,\vec P),\vec p(\vec Q,\vec P))]_{M_{(q,p)}} \not=
[G(\vec q(\vec Q,\vec P),\vec p(\vec Q,\vec P)),F(\vec q(\vec Q,\vec P),\vec p(\vec Q,\vec P))]_{M_{(Q,P)}},   
\end{equation}
which is a consequence of the fact that the star product (sometimes expressed in terms of the symmetric bracket \cite{Baker}) is also not invariant under a general coordinate transformation:
\begin{equation}
G(\vec q(\vec Q,\vec P),\vec p(\vec Q,\vec P))*_{(q,p)} F(\vec q(\vec Q,\vec P),\vec p(\vec Q,\vec P)) \not=
G(\vec q(\vec Q,\vec P),\vec p(\vec Q,\vec P))*_{(Q,P)} F(\vec q(\vec Q,\vec P),\vec p(\vec Q,\vec P)).  
\end{equation}
These features of the Weyl-Wigner map and consequently of the star product are well known and have been extensively studied in the past (see for instance \cite{Flato1,Flato2}). Namely, it was proved that the set of observables invariant under general unitary transformations is the set of first order polynomials in the fundamental variables and the coordinate transformations that preserve the star product are the linear transformations.

From the previous analysis we see that the behavior of Moyal quantum mechanics under the action of canonical and coordinate transformations is a direct consequence of the definition of the Weyl-Wigner map (15). This motivates the purpose of the next section where we will present a generalization of the Weyl-Wigner map and use it to prove that unitary operator transformations can be implemented as coordinate transformations in the Moyal formalism. Furthermore, we will see in section 4 that the new map provides a derivation of Vey covariant quantum mechanics directly from standard operator quantum mechanics. 

\section{Generalized Weyl-Wigner map}

We start by introducing a new Weyl-Wigner map in the variables $(\vec Q,\vec P)$ (which are not required to be canonical) that copies the Weyl-Wigner map in the variables $(\vec q,\vec p)$:

\underline{{\bf Definition}:} {\bf Generalized Weyl-Wigner map}\\
Let $W_{(q,p)}$ be the standard Weyl-Wigner map (in the variables $(\vec q,\vec p)$) from the algebra of linear operators $\hat{\cal A}({\cal H})$ acting on the physical Hilbert space ${\cal H}$ to the algebra of observables in the phase space $T^{\ast}M$. For the variables $(\vec Q,\vec P)$, assumed in one-to-one correspondence with the canonical variables $(\vec q,\vec p)$, we define a new {\it generalized} Weyl-Wigner map:
\begin{equation}
W_{(Q,P)}^{\prime}: \hat{\cal A}({\cal H}) \longrightarrow {\cal A}(T^{\ast}M); \quad
W_{(Q,P)}^{\prime}(\hat A)=W_{(q,p)}(\hat A) ,\quad \forall_{\hat A \in \hat{\cal A}({\cal H})}. 
\end{equation}
For each new choice of the canonical variables $(\vec q,\vec p)$ we obtain a new Weyl-Wigner map in the variables $(\vec Q,\vec P)$. If the transformation from $(\vec q,\vec p)$ to $(\vec Q,\vec P)$ is a polynomial of first degree then $W_{(Q,P)}^{\prime}=W_{(q,p)}$. Otherwise the two maps differ.\\

The aim of this section is to obtain the explicit expression for $W_{(Q,P)}^{\prime}$. We start by deriving the generalizations of the Weyl order and Weyl symbol prescriptions.

A generic dynamical operator $\hat A$ can be cast in a fully symmetrized form according to Weyl's prescription:
\begin{equation}
\hat A_{W(q,p)}=\hat A (\hat{\vec q},\hat{\vec p}) = \int d^N \vec x d^N \vec y \hspace{0.25 cm} \alpha 
( \vec x, \vec y) e^{i \vec x \cdot \hat{\vec q} + i \vec y \cdot \hat{\vec p}},  
\end{equation}
where the subscript $W(q,p)$ means that $\hat A$ is displayed as a fully symmetrized functional of the variables $(\hat{\vec q},\hat{\vec p})$, and $\vec x \cdot \hat{\vec q} + \vec y \cdot \hat{\vec p} \equiv \sum_{j=1}^N \left( x_j \hat q_j + y_j \hat p_j \right)$. 
If $\hat A$ is 
hermitean, then the numerical (usually singular) function $\alpha( \vec x, \vec y)$ is subject to the constraint 
$\alpha^* ( \vec x, \vec y) =\alpha( - \vec x, - \vec y)$. The Weyl symbol, $W_{(q,p)}(\hat A)$, associated with the operator $\hat A$ in 
eq.(25) is the c-function of $2N$ phase space variables $(\vec q, \vec p)$ given by:
\begin{equation}
A (\vec q, \vec p) \equiv W_{(q,p)} \left[ \hat A (\hat{\vec q}, \hat{\vec p} ) \right] = \int d^N \vec x d^N \vec y \hspace{0.25 cm} \alpha 
( \vec x, \vec y) e^{i \vec x \cdot \vec q + i \vec y \cdot \vec p}.
\end{equation}
Let now $(\hat{\vec Q}, \hat{\vec P} )$ be another complete set of variables in one-to-one correspondence with $(\hat{\vec q}, \hat{\vec p})$ and let us display the variables $(\hat{\vec q}, \hat{\vec p} )$ in a completely symmetrized order in the basis $(\hat{\vec Q}, \hat{\vec P})$, i.e.:
\begin{equation}
\left\{
\begin{array}{l}
\hat{\vec q} \equiv \int d^N \vec z d^N \vec w \hspace{0.25 cm} \rho_q ( \vec z, \vec w) e^{i \vec z \cdot \hat{\vec Q} + i \vec w \cdot \hat{\vec P}},\\
\\
\hat{\vec p} \equiv \int d^N \vec x d^N \vec y \hspace{0.25 cm} \rho_p ( \vec x, \vec y) e^{i \vec x \cdot \hat{\vec Q} + i \vec y \cdot \hat{\vec P}}.
\end{array}
\right.
\end{equation}
The new variables are not necessarily canonical: in general, $\left[\hat Q_i , \hat P_j \right] \ne i \hbar \delta_{ij}$. The $(\vec Q,\vec P)$-Weyl symbols associated with (27) are:
\begin{equation}
\left\{
\begin{array}{l}
\vec q \equiv W_{(Q,P)} ( \hat{\vec q}) = \int d^N \vec z d^N \vec w \hspace{0.25 cm} \rho_q ( \vec z, \vec w) e^{i \vec z \cdot \vec Q + i \vec w \cdot 
\vec P},\\
\\
\vec p \equiv W_{(Q,P)} ( \hat{\vec p}) = \int d^N \vec x d^N \vec y \hspace{0.25 cm} \rho_p ( \vec x, \vec y) e^{i \vec x \cdot \vec Q + i \vec y \cdot 
\vec P}.
\end{array}
\right.
\end{equation}
Let us now consider again the operator $\hat A$ given by eq.(25). In the $(\hat{\vec Q},\hat{\vec P})$ representation $\hat A$ is written as: $\hat A=
\hat A(\hat{\vec q}(\hat{\vec Q},\hat{\vec P}),
\hat{\vec p}(\hat{\vec Q},\hat{\vec P}))=
\hat A'(\hat{\vec Q},\hat{\vec P})$ and the explicit functional form of $\hat A'$ is given by the {\it generalized Weyl prescription}:
\begin{equation}
\hat A_{W'(Q,P)} =\hat A'(\hat{\vec Q},\hat{\vec P}) = \int d^N \vec x d^N \vec y \hspace{0.25 cm} \alpha ( \vec x, \vec y) e^{i \vec x \cdot \hat{\vec q}(\hat{\vec Q},\hat{\vec P})
 + i \vec y \cdot 
\hat{\vec p}(\hat{\vec Q},\hat{\vec P})},
\end{equation}
where $\hat{\vec q}(\hat{\vec Q},\hat{\vec P})$ and $\hat{\vec p}(\hat{\vec Q},\hat{\vec P})$  are given (in a fully symmetrized order) by eq.(27) and
the subscript $W'(Q,P)$ means that $\hat A$ is displayed as a functional of the variables $(\hat{\vec Q},\hat{\vec P})$ in a fully symmetrized order in the variables $(\hat{\vec q},\hat{\vec p})$. Notice that the numerical function $\alpha ( \vec x, \vec y)$ is the same as in eq.(25). The standard $(\vec q,\vec p)$-Weyl symbol associated with $\hat A$ is given by eq.(26) and thus, using the definition (24), it is straightfoward to conclude that the  generalized Weyl symbol associated with $\hat A$ is given by:
\begin{equation}
A'( \vec Q, \vec P ) \equiv W'_{(Q,P)} \left[ \hat A' ( \hat{\vec Q} , \hat{\vec P})
\right] =   \int d^N \vec x d^N \vec y \hspace{0.25 cm} \alpha ( \vec x, \vec y) e^{i \vec x \cdot \vec q (\vec Q, \vec P )+ i \vec y \cdot \vec p (\vec Q, \vec P )},
\end{equation}
and one immediately realizes that $A'(\vec Q, \vec P )=A( \vec q(\vec Q,\vec P), \vec p(\vec Q,\vec P) )$ as it should.

Finally, we want to derive the covariant generalization of the Weyl-Wigner map given by eq.(1). We start by re-writing $W_{(q,p)}$ as follows:
\begin{equation}
A(\vec q,\vec p)= W_{(q,p)}(\hat A)= \int d^N\vec x \int d^N\vec y e^{-i\vec p \cdot \vec y}
\delta (\vec x-\vec q) F(\vec x,\vec y), 
\end{equation}
where $F(\vec x,\vec y)=<\vec x-\frac{\hbar}{2} \vec y|\hat A|\vec x+\frac{\hbar}{2} \vec y>$ 
and $|\vec x \pm \frac{\hbar}{2} \vec y>$ are eigenstates of $\hat{\vec q}$. 
The function $F(\vec x,\vec y)$ is invariant under change of representation:
\begin{equation}
F(\vec x,\vec y)=<\vec x-\frac{\hbar}{2} \vec y|\hat A( \hat{\vec q} , \hat{\vec p})
|\vec x+\frac{\hbar}{2} \vec y> = 
<\vec x-\frac{\hbar}{2} \vec y|_Q\hat A( \hat{\vec q}( \hat{\vec Q} , \hat{\vec P}) , \hat{\vec p}( \hat{\vec Q} , \hat{\vec P}))|\vec x+\frac{\hbar}{2} \vec y>_Q ,
\end{equation}
where the subscript "$Q$" makes it explicit that the eigenstates of $\hat{\vec q}$ are displayed in the $\hat{\vec Q}$ representation. 
Furthermore it is trivial to realize that:
\begin{equation}
A(\vec q(\vec Q,\vec P),\vec p(\vec Q,\vec P))= \int d^N\vec x \int d^N\vec y e^{-i\vec p(\vec Q,\vec P) \cdot \vec y}
\delta (\vec x-\vec q(\vec Q,\vec P)) F(\vec x,\vec y) ,
\end{equation}
and thus we get the explicit expression for the covariant Weyl-Wigner map:
\begin{equation}
W'_{(Q,P)}(\hat A)= \int d^N\vec x \int d^N\vec y e^{-i\vec p(\vec Q,\vec P) \cdot \vec y}
\delta (\vec x-\vec q(\vec Q,\vec P)) 
<\vec x-\frac{\hbar}{2} \vec y|_Q \hat A |\vec x+\frac{\hbar}{2} \vec y>_Q 
\end{equation}
satisfying definition (24).

To illustrate the features of the new map let us consider the unitary transformation (14).
The generalized map yields a different phase space version of this transformation. The analogous of eq.(17) is given by:
\begin{equation}
T:A(\vec q,\vec p) = W_{(q,p)} (\hat A(\hat{\vec q},\hat{\vec p})) \longrightarrow  A^{\prime}(\vec Q,\vec P) =   
W'_{(Q,P)} (\hat{A}'(\hat{\vec Q},\hat{\vec P})),  
\end{equation}
and we get:
\begin{eqnarray}
&& W'_{(Q,P)} (\hat{A}'(\hat{\vec Q},\hat{\vec P}))=
W'_{(Q,P)} (\hat U \hat{A}(\hat{\vec Q},\hat{\vec P}) \hat U^{-1}) \nonumber \\
&=& \int d^N\vec x \int d^N\vec y e^{-i\vec p(\vec Q,\vec P) \cdot \vec y}
\delta (\vec x-\vec q(\vec Q,\vec P)) 
<\vec x-\frac{\hbar}{2} \vec y| \hat U^{-1}(\hat U \hat{A}(\hat{\vec Q},\hat{\vec P}) \hat U^{-1}) \hat U |\vec x+\frac{\hbar}{2} \vec y> \nonumber \\
& = & A(\vec q(\vec Q,\vec P),\vec p(\vec Q,\vec P)) ,
\end{eqnarray}
where $\vec q(\vec Q,\vec P)$ and $\vec p(\vec Q,\vec P)$ are given by eq.(16) and
$|\vec x \pm \frac{\hbar}{2} \vec y>$ are eigenstates of $\hat{\vec Q}$ and thus 
$ \hat U |\vec x \pm \frac{\hbar}{2} \vec y>$ are eigenstates of $\hat{\vec q}$, displayed in the $\hat{\vec Q}$ representation, and
 with associated eigenvalues $\vec x \pm \frac{\hbar}{2} \vec y$.
As expected, the previouse result means that the unitary transformation is mapped by the generalized Weyl-Wigner map to a phase space coordinate transformation.

\section{Covariant star product}

The new Weyl-Wigner map yields a new star product through the definition:
\begin{equation}
W^{\prime}_{(Q,P)} (\hat A \cdot \hat B) = W^{\prime}_{(Q,P)} (\hat A) *^{\prime}_{(Q,P)} 
W^{\prime}_{(Q,P)} (\hat B); \quad \forall_{\hat A, \hat B \in \hat{\cal A}({\cal H})},
\end{equation}
and one immediately recognizes that the new product satisfies:
\begin{eqnarray}
&& W^{\prime}_{(Q,P)} (\hat A) *^{\prime}_{(Q,P)} 
W^{\prime}_{(Q,P)} (\hat B)= W^{\prime}_{(Q,P)} (\hat A \hat B) \nonumber \\
&=& W_{(q,p)} (\hat A \hat B) = W_{(q,p)} (\hat A) *_{(q,p)} 
W_{(q,p)} (\hat B).
\end{eqnarray}
If $W_{(q,p)} (\hat A)=A(\vec q,\vec p)$ then $W'_{(Q,P)} (\hat A)=A(\vec q(\vec Q,\vec P),\vec p(\vec Q,\vec P))=A'(\vec Q,\vec P)$ and thus:
\begin{equation}
A'(\vec Q,\vec P) *^{\prime}_{(Q,P)} B'(\vec Q,\vec P) = A(\vec q(\vec Q,\vec P),\vec p(\vec Q,\vec P))*_{(q,p)} B(\vec q(\vec Q,\vec P),\vec p(\vec Q,\vec P)) \qquad \forall_{A,B \in {\cal A}(T^{\ast}M)}.
\end{equation}
The former result immediately implies that the new product is also a non-commutative, associative product for the algebra of functions over the classical phase space. Moreover, using the new product we can define a new bracket (named generalized Moyal bracket) alternative to the standard Moyal bracket:
\begin{equation}
[A,B]_{M^{\prime}_{(Q,P)}}=A *^{\prime}_{(Q,P)} B - B *^{\prime}_{(Q,P)} A.
\end{equation}
It follows from (39) that this is also a Lie bracket.

Neither the new star product nor the new Moyal bracket display the same functional structure as the standard ones $*_{(Q,P)}$ and $[,]_{M_{(Q,P)}}$. The aim of the rest of this section is to derive the explicit form of the new product and in the sequel of the new bracket from the generalized Weyl-Wigner map.
Let us start by introducing the notation:
$$
\left\{
\begin{array}{l l}
\hat O^k = \hat p_k, & k=1, \cdots ,N\\
\hat O^k = \hat q_{k-N}, & k=N+1, \cdots, 2N
\end{array}
\right.
\qquad 
\left\{
\begin{array}{l l}
\hat O'^k = \hat P_k, & k=1, \cdots ,N\\
\hat O'^k = \hat Q_{k-N}, & k=N+1, \cdots, 2N
\end{array}
\right.
$$
and the symbols $O^k \equiv W_{(q,p)} (\hat O^k)$, $O'^k \equiv W_{(Q,P)} (\hat O'^k)$. Consider the two following operators displayed in the generalized Weyl order\footnote{Sum over repeated indices is understood.}:
\begin{equation}
\left\{
\begin{array}{l}
\hat A_{W'(Q,P)}= \hat A' ( \hat{\vec Q}, \hat{\vec P}) = \int d^{2N} \vec a \hspace{0.25 cm} \alpha ( \vec a) e^{ i a_k \hat O^k(\hat O'^s)} ,\\
\\
\hat B_{W'(Q,P)}=\hat B' ( \hat{\vec Q}, \hat{\vec P}) = \int d^{2N} \vec b \hspace{0.25 cm} \beta ( \vec b) e^{ i b_l \hat O^l(\hat O'^r)} ,
\end{array}
\right.
\end{equation}
where $\vec a =(a_1, \cdots , a_{2N})$, $\vec b =(b_1, \cdots , b_{2N})$. Let us then calculate $W^{\prime}_{(Q,P)} (\hat A \cdot \hat B)$ explicitly and express the result in terms of the symbols $A'({\vec Q},{\vec P})=W^{\prime}_{(Q,P)}(\hat A)$ and $B'({\vec Q},{\vec P})=W^{\prime}_{(Q,P)}(\hat B)$:
\begin{equation}
\begin{array}{c}
W^{\prime}_{(Q,P)}(\hat A \cdot \hat B) = \int d^{2N} \vec a d^{2N} \vec b \hspace{0.25 cm} \alpha (\vec a) \beta (\vec b) W^{\prime}_{(Q,P)} \left\{  e^{i a_k \hat O^k(\hat O'^s)}
 \cdot  e^{i b_l \hat O^l(\hat O'^r)}  \right\}=\\
\\
= \int d^{2N} \vec a d^{2N} \vec b \hspace{0.25 cm} \alpha (\vec a) \beta (\vec b) W_{(q,p)} \left\{  e^{i a_k \hat O^k(O'^s)}
 \cdot  e^{i b_l \hat O^l(O'^r)}  \right\}=\\
\\
=\int d^{2N} \vec a d^{2N} \vec b \hspace{0.25 cm} \alpha (\vec a) \beta (\vec b) \left [ e^{i a_k O^k(O'^s)}
\right] *_{(q,p)} \left [ e^{i b_l  O^l(O'^r)} \right].
\end{array}
\end{equation}
Using the explicit expression of the $(\vec q,\vec p)$ star product eq.(5) we obtain the following expansion in powers of $\hbar$:
\begin{equation}
W'_{(Q,P)} (\hat A \cdot \hat B) = \sum_{k=0}^{\infty} \frac{1}{k!} \left(\frac{i \hbar}{2} \right)^k  
\int d^{2N} \vec a d^{2N} \vec b \hspace{0.25 cm} \alpha (\vec a) \beta (\vec b) \left [ e^{i a_t O^t (O'^s)}
\right] \hat{\cal J}_k \left [ e^{i b_l  O^l(O'^r)} \right],
\end{equation}
where
\begin{equation}
\hat{\cal J}_k=
{\buildrel { \leftarrow}\over\partial}_{i_1} ...{\buildrel { \leftarrow}\over\partial}_{i_k} 
J_{(q,p)}^{{i_1}{j_1}}...J_{(q,p)}^{{i_k}{j_k}} 
{\buildrel { \rightarrow}\over\partial}_{j_1}...{\buildrel { \rightarrow}\over\partial}_{j_k}.
\end{equation}
At this point we recall that the phase space is assumed to have the structure of a flat manifold and introduce the $2N \times 2N$ "Euclidean" metric\footnote{$\alpha$ and $\beta$ are arbitrary constants introduced to ensure the correct dimensions.}:
\begin{equation}
\left( g_{ij} \right) = \left(
\begin{array}{l r}
\alpha 1_{N \times N} & 0_{N \times N}\\
0_{N \times N} & \beta 1_{N \times N}
\end{array}
\right),
\end{equation}
and the associated covariant derivative $\nabla_i$:
\begin{equation}
\begin{array}{l l}
\nabla_i A  & = \partial_i A,\\
\nabla_i \nabla_j A & = \partial_i \partial_j A - \Gamma^k_{ij} \partial_k A, \qquad i,j,k= 1, \cdots, 2N,
\end{array}
\end{equation}
where the Christoffel symbols $\Gamma^k_{ij}$ are fully determined by the metric:
\begin{equation}
\Gamma^i_{jk} = \frac{1}{2} g^{il} \left( \partial_k g_{lj} + \partial_j g_{lk} - \partial_l g_{jk} \right), \qquad i,j,k =
1, \cdots , 2N.
\end{equation}
Obviously, in the coordinates $(\vec q, \vec p)$ we have
$\Gamma^i_{jk} =0, \quad  \forall i,j,k =1, \cdots , 2N$
and thus $\nabla_i=\partial_i$.

Under the general coordinate transformation $O^i \to O^i(O'^s)$, $(i=1, \cdots, 2N)$, the symplectic matrix and the covariant derivative transform according to: 
\begin{equation}
J'^{ij}_{(Q,P)} = \frac{\partial O'^i}{\partial O^k} \frac{\partial O'^j}{\partial O^l} J^{kl}_{(q,p)} = \left\{ O'^i, O'^j \right\}_{(q,p)}  = O'^i {\hat{\cal J}}_{(q,p)} O'^j,
\end{equation}
\begin{equation}
\Gamma'^i_{jk} = \Gamma^m_{bc} \frac{\partial O'^i}{\partial O^m} \frac{\partial O^b}{\partial O'^j} \frac{\partial
O^c}{\partial O'^k} + \frac{\partial O'^i}{\partial O^b} \frac{\partial^2 O^b}{\partial O'^j \partial O'^k} = 
 \frac{\partial O'^i}{\partial O^b} \frac{\partial^2 O^b}{\partial O'^j \partial O'^k}.
\end{equation}
Moreover the terms $ \left [ e^{i a_t O^t (O'^s)}
\right] \hat{\cal J}_k \left [ e^{i b_l  O^l(O'^r)} \right]$ are scalars and thus are left invariant: 
\begin{eqnarray}
&& \left [ e^{i a_t O^t (O'^s)}\right] 
{\buildrel { \leftarrow}\over\partial}_{i_1} ...{\buildrel { \leftarrow}\over\partial}_{i_k} 
J_{(q,p)}^{{i_1}{j_1}}...J_{(q,p)}^{{i_k}{j_k}} 
{\buildrel { \rightarrow}\over\partial}_{j_1}...{\buildrel { \rightarrow}\over\partial}_{j_k}
 \left [ e^{i b_l  O^l(O'^r)} \right] \nonumber \\
&=&
\left [ e^{i a_t O^t(O'^s)}\right] 
{\buildrel { \leftarrow}\over\nabla'}_{i_1} ...{\buildrel { \leftarrow}\over\nabla'}_{i_k} 
J'^{{i_1}{j_1}}_{(Q,P)}...J'^{{i_k}{j_k}}_{(Q,P)} 
{\buildrel { \rightarrow}\over\nabla'}_{j_1}...{\buildrel { \rightarrow}\over\nabla'}_{j_k}
 \left [ e^{i b_l  O^l(O'^r)} \right],
\end{eqnarray}
where the new covariant derivative $\nabla'$ is given by:
\begin{equation}
\begin{array}{l l}
\nabla'_i A  & = \partial'_i A,\\
\nabla'_i \nabla'_j A & = \partial'_i \partial'_j A - \Gamma'^k_{ij} \partial'_k A, \quad \partial'_i= \partial /\partial {O'}^{i};
\qquad i,j,k= 1, \cdots, 2N.
\end{array}
\end{equation}
Substituting the result (50) in eq.(43) and taking into account (41), it is trivial to obtain: $W'_{(Q,P)} (\hat A \cdot \hat B) =
W'_{(Q,P)} (\hat A) *'_{(Q,P)} W'_{(Q,P)} ( \hat B)$ where the new star product is given by: 
\begin{equation}
A'( \vec Q, \vec P) *'_{(Q,P)} B'( \vec Q, \vec P)= A'( \vec Q, \vec P) e^{\frac{i \hbar}{2} {\buildrel { \leftarrow}\over\nabla'}_i  J'^{ij}_{(Q,P)}  {\buildrel 
{ \rightarrow}\over\nabla'}_j} B' (\vec Q, \vec P),
\end{equation} 
and we recovered the covariant formulation of the star product first introduced by Vey \cite{Vey}. The covariant formulation ensured the invariant nature of the numerical value for the star-product of two observables in any coordinate system. However, in
general, the functional form of the product is altered under an arbitrary coordinate transformation. 

Finally, we can easily obtain the functional form of the new bracket: $
\left[A', B' \right]_{M'_{(Q,P)}} =
W_{(Q,P)}^{\prime}([\hat A,\hat B])$:
\begin{equation}
\left[ A'(\vec Q,\vec P), B'(\vec Q,\vec P) \right]_{M'_{(Q,P)}} = \frac{2}{\hbar} A'(\vec Q,\vec P) \sin \left(\frac{\hbar}{2} 
{\buildrel { \leftarrow}\over\nabla'}_i  J'^{ij}_{(Q,P)}  {\buildrel 
{ \rightarrow}\over\nabla'}_j \right) B'(\vec Q,\vec P),
\end{equation}
and if $(\vec Q,\vec P)$ is a set of canonical variables this is equally a deformation of the Poisson bracket:
\begin{equation}
\left[A', B' \right]_{M'_{(Q,P)}} \equiv \frac{1}{ i \hbar} \left( A' *'_{(Q,P)} B' - B' *'_{(Q,P)} A' \right) = \left\{ A', B' \right\}_{(Q,P) } + O({\hbar}^2).
\end{equation}

\section{Covariant Vey quantum mechanics and discussion}

The covariant formulation of Moyal quantum mechanics lives on the classical phase space $T^{\ast}M$ with the structure of the tangent bundle of the configuration space, where a symplectic structure $J^{ij}$ and a metric structure $g_{ij}$ (or alternatively a Poisson connection) can be naturally defined. 

The fundamental mathematical objects of the theory are the Wigner distribution function $f'_W(\vec Q,\vec P;t)$ 
and the observables $A'(\vec Q,\vec P)$. They all are scalar functions over $(T^{\ast}M,J^{ij},g_{ij})$, and are related with the mathematical objects of standard operator quantum mechanics through the generalized Weyl-Wigner map: $f'_W(\vec Q,\vec P;t)=W'_{(Q,P)}(|\psi(t)><\psi(t)|)$ and $A'(\vec Q,\vec P;t)=W'_{(Q,P)}(\hat A)$
The time evolution of the Wigner function is given by the dynamical equation:
\begin{equation}
\dot{f'}_W=[H',f'_W]_{M'_{(Q,P)}}
\end{equation}
which transforms covariantly under arbitrary phase space difeomorphisms yielding, in any coordinates, identical mathematical solutions and thus identical physical predictions.
 
Finally, the covariant form of eqs. (7) and (8) yield the physical relevant predictions. The Moyal star-genvalue equation can be obtained, through the Weyl-Wigner map, from the standard operator eigenvalue equation: $\hat A |\psi^n_a><\psi^n_a|=a |\psi^n_a><\psi^n_a|$ (where $n$ is the degeneracy index). We follow the same procedure but use the generalized Weyl-Wigner map and obtain:
\begin{equation}
A'(\vec Q,\vec P) *'_{(Q,P)} g'^n_a(\vec Q,\vec P)= a g'^n_a(\vec Q,\vec P).
\end{equation}
It is trivial to check that $g'^n_a(\vec Q,\vec P)=g^n_a(\vec q(\vec Q,\vec P),\vec p(\vec Q,\vec P))$ where $g^n_a(\vec q,\vec p)$ is the solution of the original star-genvalue equation.
Futhermore, the probabilistic functionals are just the coordinate transform of the original ones: 
\begin{equation}
P(A'(\vec Q,\vec P;t)=a)= \sum_n \int d^N\vec Q \int d^N\vec P (\mbox{det} J'^{ij}_{(Q,P)})^{-1/2} 
g'^n_a(\vec Q,\vec P) f'_W(\vec Q,\vec P;t),
\end{equation}
and finally the average value prediction is also trivially covariantized:
\begin{equation}
<A'(\vec Q,\vec P;t)>= \int d^N\vec Q \int d^N\vec P (\mbox{det} J'^{ij}_{(Q,P)})^{-1/2} A'(\vec Q,\vec P) f'_W(\vec Q,\vec P;t).
\end{equation}

Let us then summarize our results: we presented an original generalized (covariant) formulation of the Weyl-order operator prescription and of the Weyl-Wigner map yielding the entire structure of Vey quantum mechanics directly from standard operator quantum mechanics. All key ingredients of the theory were derived in this fashion thus casting Vey quantum mechanics at the same level of completeness as Moyal quantum mechanics. Furthermore we studied the action of standard operator transformations in the Moyal formalism and concluded that through the generalized Weyl-Wigner map these transformations can be implemented as phase space coordinate transformations in the Moyal idiom.
It is now easy to realize that the group of "canonical" transformations of Vey quantum mechanics (those that preserve the bracket structure) is the subgroup of the symplectic transformations which are also isometries, i.e. the coordinate transformations $O^i \to O^i(O'^s)$ such that:
\begin{equation}
\begin{array}{l}
J'^{ij}_{(Q,P)} = \frac{\partial O'^i}{\partial O^k} \frac{\partial O'^j}{\partial O^l} J^{kl}_{(q,p)} = J^{ij}_{(q,p)},\\
\\
\Gamma'^i_{jk} = \frac{\partial O'^i}{\partial O^b} \frac{\partial^2 O^b}{\partial O'^j \partial O'^k}=0,
\end{array}
\end{equation}
and notice that in the limit $\hbar \to 0$ the requirement on the isometry character of the canonical transformations disappears, as it should, and we recover the standard sympletic group of classical mechanics.

Finally, let us make a small remark concerning possible future applications of the generalized Weyl-Wigner map. Recently there has been a considerable interest in developing a fully classical interpretation for Moyal dynamics \cite{Flato1,Flato2,nuno3}. This is a difficult task for several reasons of which the most important are that Moyal dynamics is non-local and there is no general criterion
to understand what is meant by "classical". Very recently, however the authors presented a proposal for a classicality criterion and proved that it is fully compatible with the Weyl-Wigner map \cite{nuno1,nuno3}. One of the problems that were left unsolved was how to understand the highly non-classical behavior of an allegedly Moyal classical dynamics, under the action of canonical transformations. This problem has been previously considered in \cite{Flato1,Flato2} using an approach based on a privileged set of observables. 

We may now expect that if the results of \cite{nuno3} (concerning the compatibility between the classicality criteria and the Weyl-Wigner map) turn out to be extendable to the covariant Weyl-Wigner map, then they will most likely open the path for a consistent classical interpretation of the action of canonical transformations in Moyal dynamics. This will be the subject of a future work \cite{nuno7}.

\section{Example}

To illustrate some of the features of the formalism let  
us consider the system of two interacting particles described by the Hamiltonian:
\begin{equation}
\hat H=\frac{\hat p^2}{2M} + \frac{\hat y^2}{2m} +k\hat q\hat y^2,
\end{equation}
where $(\hat q,\hat p)$ are the fundamental variables of the particle of mass $M$, $(\hat x,\hat y)$ the ones of the particle of mass $m$ and $k$ is a coupling constant.\\

{\it a) Standard description in the original variables}. In the Heisenberg picture the time evolution of the former system is given by:
\begin{equation}
\left\{ \begin{array}{l}
\hat q(t)=\hat q+\frac{\hat p}{M}t-\frac{k}{2M}\hat y^2t^2 \\
\hat p(t)=\hat p-k\hat y^2t \\
\hat x(t)=\hat x+\{ \frac{\hat y}{m}+2k\hat q\hat y\} t+\frac{k}{M}\hat p\hat yt^2 - \frac{k^2}{3M}\hat y^3t^3 \\
\hat y(t)=\hat y 
\end{array}  \right. 
\end{equation}
The standard Weyl-Wigner transform yields the Moyal description of the system. The Hamiltonian is trivially obtained:
\begin{equation}
H = W_{(q,x,p,y)}(\hat H)=\frac{p^2}{2M} + \frac{y^2}{2m} +kqy^2,
\end{equation}
and the Moyal time evolution of the system is given by the dynamical equations:
\begin{equation}
\dot A = [A,H]_{M_{q,x,p,y}} ,
\end{equation}
which yield the following solutions (to make it simpler let us concentrate on the dynamics of the particle of mass $M$):
\begin{equation}
\left\{ \begin{array}{l}
q(t)=q(0)+\frac{p(0)}{M}t-\frac{k}{2M}y(0)^2t^2 \\
p(t)=p(0)-ky(0)^2t 
\end{array}  \right. 
\end{equation}
Notice that the same predictions can be obtained by applying the Weyl-Wigner map to the solutions (61).\\

{\it b) Canonical transformation and the standard description in the new variables}. Let us now consider the canonical transformation:
\begin{equation}
\left\{ \begin{array}{l}
\hat q = \hat q \\
\hat p = \hat p 
\end{array}  \right. \quad , \quad
\left\{ \begin{array}{l}
\hat x = \ln \hat Q \\
\hat y = \frac{1}{2} (\hat Q \hat P + \hat P \hat Q). 
\end{array}  \right. 
\end{equation}
In the new variables $\hat H$ takes the form:
\begin{equation}
\hat H=\frac{\hat p^2}{2M} + \frac{(\hat Q \hat P + \hat P \hat Q)^2}{8m} +\frac{k}{4}\hat q 
(\hat Q \hat P + \hat P \hat Q)^2,
\end{equation}
and it yields the Heisenberg picture time evolution (still just for the particle of mass $M$):
\begin{equation}
\left\{ \begin{array}{l}
\hat q(t)=\hat q+\frac{\hat p}{M}t-\frac{k}{8M}(\hat Q \hat P + \hat P \hat Q)^2t^2 \\
\hat p(t)=\hat p-\frac{k}{4}(\hat Q \hat P + \hat P \hat Q)^2t 
\end{array}  \right. 
\end{equation}
To obtain the Moyal formulation of the system in the variables $(q,Q,p,P)$ the first step is to use the standard Weyl-Wigner map to get the phase space representation of the transformation (65):
\begin{equation}
\left\{ \begin{array}{l}
q = W_{(q,Q,p,P)} (\hat q)=q \\
p = W_{(q,Q,p,P)} (\hat p)=p 
\end{array}  \right. \quad , \quad
\left\{ \begin{array}{l}
x = W_{(q,Q,p,P)} (\hat x(\hat Q,\hat P)) = \ln Q \\
y = W_{(q,Q,p,P)} (\hat y(\hat Q,\hat P))=  Q P  
\end{array}  \right. 
\end{equation}
We then express the Hamiltonian using the standard Weyl order prescription: 
\begin{equation}
\hat H_{W(q,Q,p,P)}=\frac{\hat p^2}{2M} + \frac{(\hat P^2 \hat Q^2)_S}{2m} +k\hat q 
(\hat Q^2 \hat P^2)_S + \frac{k}{4} \hbar^2 \hat q + \frac{\hbar^2}{8m},
\end{equation}
where we used the fact that $\frac{1}{4}(\hat Q \hat P + \hat P \hat Q)^2_{W(q,Q,p,P)}=
(\hat P^2 \hat Q^2)_S + \hbar^2/4$, (the subscripted "$S$" standing for the full symmetrization of the operator) and so $\frac{1}{4}(\hat Q \hat P + \hat P \hat Q)^2$ can be expressed in the standard Weyl order (eq.(25)) by making $\alpha(a,b,c,d)= \delta(a)\delta(b)\{ \delta''(c)\delta''(d)+\hbar/4 \delta (c) \delta(d)\}$ where $a,b,c,d$ are the integration variables associated to the fundamental operatores $\hat q,\hat p,\hat Q,\hat P$, respectively. The standard Weyl-Wigner transform of $\hat H$ is thus:
\begin{equation}
H= W_{(q,Q,p,P)}(\hat H)=\frac{p^2}{2M} + \frac{P^2 Q^2}{2m} +k q 
Q^2 P^2 + \frac{k}{4} \hbar^2 q + \frac{\hbar^2}{8m}.
\end{equation}
Finally, the Moyal dynamical equations in the variables $(q,Q,p,P)$ yield:
\begin{equation}
\left\{ \begin{array}{l}
q(t)=q(0)+\frac{p(0)}{M}t-\frac{k}{2M}Q(0)^2P(0)^2t^2 - \frac{k\hbar^2}{8M}t^2 \\
p(t)=p(0)-kQ(0)^2P(0)^2t - \frac{k\hbar^2}{4} t^2.
\end{array}  \right. 
\end{equation}
Notice that equivalent results can be obtained by applying the Weyl-Wigner transform to the time evolution operator equations (67). Furthermore, we realize that the transformation (65) does not act as a coordinate transformation in the Moyal formalism. Taking for instance the Hamiltonian, we have: $H(q,p,x(Q,P),y(Q,P)) \not= W_{(q,Q,p,P)}(\hat H)$ the left hand side being given by eq.(62) and the right hand side by eq.(70). Consequently the two phase space orbits (eqs.(64,71)) are not the coordinate transformation of each other.\\

{\it c) Covariant formulation}. Finally, let us use the generalized Weyl-Wigner map. In the generalized Weyl order prescription eq.(29), we make $\alpha(a,b,c,d)= -\delta(a)\delta(b)\delta(c)\delta''(d)$ - where, this time $a,b,c,d$ are the integration variables associated to the fundamental operators $\hat q,\hat p,\hat x(\hat Q,\hat P),\hat y(\hat Q,\hat P)$ - and get:
\begin{equation}
\frac{1}{4}(\hat Q \hat P + \hat P \hat Q)^2_{W'(q,Q,p,P)}=
\frac{1}{4}(\hat Q \hat P \hat Q \hat P + \hat Q \hat P^2 \hat Q + \hat P \hat Q^2 \hat P + \hat P \hat Q \hat P \hat Q ),
\end{equation}
and thus:
\begin{equation}
H'(q,Q,p,P)= W'_{(q,Q,p,P)}(\hat H)=\frac{p^2}{2M} + \frac{Q^2 P^2}{2m} +k q 
Q^2 P^2, 
\end{equation}
which, as expected is the coordinate transformation of the observable $H$ given by eq.(62). The new star product and the new bracket are given by eqs.(52) and (53), respectively, where $J'^{ij}_{(Q,P)}=J^{ij}_{(q,p)}$ and the covariant derivatives are associated to the Christoffel symbols:
\begin{equation}
\Gamma'^1_{11} = -1/Q; \quad \Gamma'^2_{11} = P/Q^2; \quad \Gamma'^2_{12} =\Gamma'^2_{21} =
1/Q,
\end{equation} 
all the others being zero (we used the notation: $O^1=x; O^2=y; O'^1=Q; O'^2=P; O'^3=O^3=q; O'^4=O^4=p$). Notice that using the new product we can go back and obtain the Hamiltonian $H'$ through a slightly different procedure, by making:
\begin{eqnarray}
&& W'_{(q,Q,p,P)}\left\{\frac{1}{2}(\hat Q \hat P + \hat P \hat Q)
\frac{1}{2}(\hat Q \hat P + \hat P \hat Q)\right\} \\
& =&  W'_{(q,Q,p,P)}\left\{\frac{1}{2}(\hat Q \hat P + \hat P \hat Q)\right\}
*'_{(q,Q,p,P)}
W'_{(q,Q,p,P)}\left\{\frac{1}{2}(\hat Q \hat P + \hat P \hat Q)\right\}
=QP *'_{(q,Q,p,P)} QP = Q^2P^2, \nonumber
\end{eqnarray}
where in the last step we used the explicit expression for the new star product (52,74).
It is now easy to check that the generalized Moyal dynamical equations,
$\dot{O}'^{i}= [O'^{i},H']_{M'_{(q,Q,p,P)}}$
yield solutions that are just the coordinate transformation of the original ones (64). Equivalent results can obviously be obtained by applying the generalized Weyl-Wigner map to the operator time evolution equations (67). 

\subsection*{Acknowledgments} 

This work was partially supported by the grants ESO/PRO/1258/98 and CERN/P/Fis/15190/1999.

\end{document}